\documentclass[9pt,conference]{IEEEtran}
\usepackage{amsmath,amssymb,amsfonts}
\usepackage{algcompatible}
\usepackage{graphicx}
\usepackage{textcomp}
\usepackage{xcolor}
\usepackage[most]{tcolorbox}

\usepackage{algpseudocode}
\usepackage{algorithm}
\usepackage{mathtools}
\usepackage[colorlinks]{hyperref}
\usepackage{indentfirst}
\usepackage[noabbrev,capitalize,nameinlink]{cleveref}
\usepackage[style=ieee,sorting=none,backend=bibtex]{biblatex}
\usepackage{subcaption}
\usepackage{multirow}
\usepackage{svg-extract}

\usepackage{tikz}
\usepackage{svg}

\usetikzlibrary{shapes.geometric, arrows}

\tikzstyle{startstop} = [rectangle, rounded corners, minimum width=3cm, minimum height=1cm,text centered,'green', 'purple' draw=black, fill=red!30]
\tikzstyle{io} = [trapezium, trapezium left angle=70, trapezium right angle=110, minimum width=3cm, minimum height=1cm, text centered, draw=black, fill=blue!30]
\tikzstyle{process} = [rectangle, minimum width=3cm, minimum height=1cm, text centered, draw=black, fill=orange!30]
\tikzstyle{decision} = [diamond, minimum width=3cm, minimum height=1cm, text centered, draw=black, fill=green!30]
\tikzstyle{arrow} = [thick,->,>=stealth]

\usepackage{xcolor}
\usepackage{url}

\title{DARIS: An Oversubscribed Spatio-Temporal Scheduler for Real-Time DNN Inference on GPUs}

\author{
    \IEEEauthorblockN{Amir Fakhim Babaei} 
    \IEEEauthorblockA{\textit{Department of Electrical \& Computer Engineering,} \\
        \textit{Virginia Tech}\\
        Arlington, Virginia, USA \\
        babaei@vt.edu\\
        \vspace{-20pt}}     
    \and    
    \IEEEauthorblockN{Thidapat Chantem} 
    \IEEEauthorblockA{\textit{Department of Electrical \& Computer Engineering,} \\
        \textit{Virginia Tech}\\
        Arlington, Virginia, USA \\
        tchantem@vt.edu\\
        \vspace{-20pt}}
}

\begin{document}
    \maketitle
    \begin{abstract}
       The widespread use of Deep Neural Networks (DNNs) is limited by high computational demands, especially in constrained environments. GPUs, though effective accelerators, often face underutilization and rely on coarse-grained scheduling. This paper introduces DARIS, a priority-based real-time DNN scheduler for GPUs, utilizing NVIDIA’s MPS and CUDA streaming for spatial sharing, and a synchronization-based staging method for temporal partitioning. In particular, DARIS improves GPU utilization and uniquely analyzes GPU concurrency by oversubscribing computing resources. It also supports zero-delay DNN migration between GPU partitions. Experiments show DARIS improves throughput by 15\% and 11.5\% over batching and state-of-the-art schedulers, respectively, even without batching. All high-priority tasks meet deadlines, with low-priority tasks having under 2\% deadline miss rate. High-priority response times are 33\% better than those of low-priority tasks.
    \end{abstract}

    \section{Introduction}
    Deep Neural Networks (DNNs) demand high computational power and challenge resource-constrained systems. GPUs improve DNN training and inference \cite{he2021enabling}, but sequential processing leads to underutilization, requiring multiple parallel (multi-tenant) DNN scheduling \cite{xu2021parallelizing}. Inference tasks in constrained settings often need real-time performance \cite{ye2024deep}, common in fields like autonomous driving \cite{liu2022prophet}, healthcare \cite{roshan2024deep, mostafaei2023ensemble}, AI at the edge \cite{aghapour2023task}, and NLP \cite{li2023rt, babaei2024hybrid}. GPU inference servers batch inputs for utilization \cite{dhakal2020gslice, guo2023optimum}, but real-time schedulers cannot typically use input batching, as they require immediate task handling \cite{yu2022survey}.

    NVIDIA provides various concurrency mechanisms for designers, but no general guideline exists on the efficient GPU configurations and resource partitioning strategies in the literature. Most works isolate resources, with some exploring oversubscription \cite{zhang2019laius}. Also, most works focus on maximizing throughput rather than achieving predictable real-time performance. GPUs have a gray-box architecture that makes timing predictions difficult. Approaches that consider \textit{Worst-Case Execution Time} (WCET) prediction often either sacrifice throughput \cite{gujarati2020serving} or result in overly pessimistic estimates \cite{wang2021balancing}.% To address this, we use a dynamic history-based approach for estimating execution times, aiming to maximize the number of deadlines met.

    Each GPU has $N_{SM,max}$ \textit{Streaming Multiprocessors} (SMs), its smallest independent units. GPU processes, or \textit{kernels}, may number in the hundreds per DNN. Kernels run sequentially in \textit{CUDA Streams} (referred to as streams after this), and running multiple streams in parallel improves GPU utilization by reducing SMs' idle time. With Multi-Process Service (MPS), we can create multiple \textit{CUDA Contexts} (referred to as contexts after this), each assigned a portion of SMs. When total SMs allocated to contexts surpass $N_{SM,max}$, it is termed \textit{Oversubscription}.

    In this paper, we propose \textit{DARIS}, a \underline{D}eadline-\underline{A}ware \underline{R}eal-Time DNN \underline{I}nference \underline{S}cheduler to address efficient GPU concurrency configuration challenges. DARIS can provide better predictability for periodic soft real-time tasks with two priority levels. Our primary goals are to minimize the deadline miss rate and maximize overall throughput. While we cannot guarantee that all high-priority tasks will meet their deadlines, DARIS demonstrates significantly fewer deadline misses compared to state-of-the-art methods. We also introduce \textit{staging} as a coarse-grained preemption mechanism to achieve better priority-based scheduling. Our main contributions are as follows:
    
    \begin{itemize}
        \item To the best of our knowledge, this is the first work to conduct an in-depth analysis of GPU concurrency mechanisms, focusing on resource oversubscription and its benefits on real platforms.
        \item We propose a deadline-aware real-time spatio-temporal GPU scheduler that surpasses the throughput of single-tenant batching, without relying on batched inputs.
    \end{itemize}

    \section{\label{sec:motive}Design Choices}
    \subsection{\label{sec:concurrency}What Concurrency Mechanism to Use?}    
    SGPRS \cite{babaei2024sgprs} defines ``knee point'' as the maximum number of DNNs a scheduler can handle without missing deadlines. They found that without temporal partitioning, throughput drops significantly beyond the knee point due to interference. Higher oversubscription values generally improve throughput, though not consistently. These results suggest that a spatio-temporal scheduler is more effective than a purely spatial one, but further analysis is needed to comprehensively examine contexts, streams, and oversubscription levels.

    \subsection{\label{sec:mot-subscribe}Is Oversubscription Good?}
    Systematic studies on SM oversubscription are limited. NVIDIA suggests 200\% SM oversubscription as optimal \cite{nvidia_mps}, though some research indicates it may cause interference \cite{dhakal2022slice, dhakal2023d, rattihalli2023fine}, with limited experimental data. Laius \cite{zhang2019laius} uses structured oversubscription, dedicating 100\% of SMs to user-facing and another 100\% to batched services. SGPRS \cite{babaei2024sgprs} highlighted the benefits of oversubscription, with the remaining challenge of finding the best trade-off. In \cref{sec:oversubscribe}, we compare oversubscription levels, denoted $OS \geq 1$, to find this balance.

    \begin{figure}
        \centering
        \includegraphics[width=1 \linewidth]{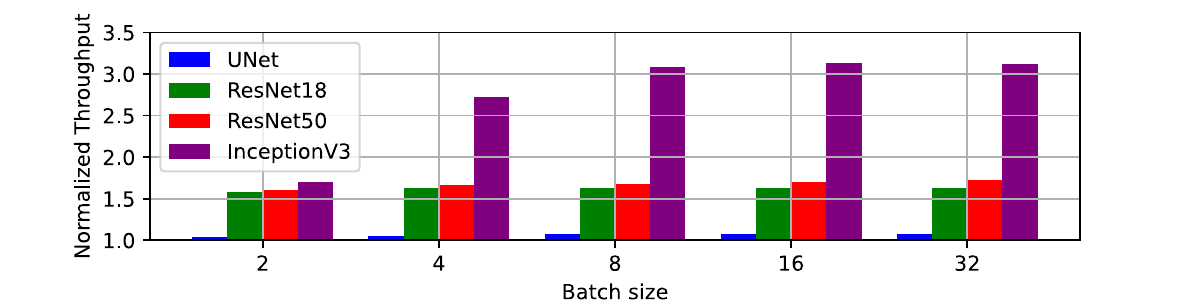}
        \caption{\label{fig:batch}Normalized throughput using batching}
    \end{figure}
    
    \subsection{Is Batching Enough or Necessary?}
    Batching boosts throughput in GPUs but is often impractical for real-time inference with non-identical DNNs since waiting for jobs can cause missed deadlines. This study examines whether batching is sufficient and if its throughput can be matched without it. Experiments (\cref{fig:batch}) show limited benefits for some DNNs (e.g., UNet) and significant gains for others (e.g., InceptionV3) with batching. DARIS achieves higher throughput without batching (\cref{sec:result}).

    \subsection{How to Predict the Execution Time of DNNs?}
    Timing prediction is essential for real-time schedulers, but GPU parallelism makes execution times unpredictable. Clockwork \cite{gujarati2020serving} ensures predictable WCET by scheduling one DNN at a time, sacrificing throughput for predictability. Multi-tenant systems face resource contention, leading to variability and overly conservative WCET estimates. Wang \textit{et al.} use history-based WCET from 10-minute simulations \cite{wang2021balancing}, also yielding conservative results. We propose a dynamic history-based prediction (\cref{fig:res_wet}) for more optimistic estimates for soft real-time systems.

    \section{\label{sec:system}System Model}
    \subsection{\label{sec:model}Task Model}
    We consider a set of $N_c$ contexts, each with $N_{SM}$ SMs. Periodic tasks correspond to individual DNNs, divided into sub-tasks or stages. We define a task set $TS=\{\tau_1, ..., \tau_{N_{ts}}\}$ with $N_{ts}$ tasks, where each task $\tau_i$ has $n_i$ sequential sub-tasks ($\tau_i=\{\tau_{i,1}, ..., \tau_{i,n_i}\}$). Tasks have two priority levels: high-priority (HP) and low-priority (LP), with counts $N_h$ and $N_l$, respectively so that $N_{ts}=N_h+N_l$.
    
\begin{figure}
        \centering
        \includegraphics[width=0.8 \linewidth]{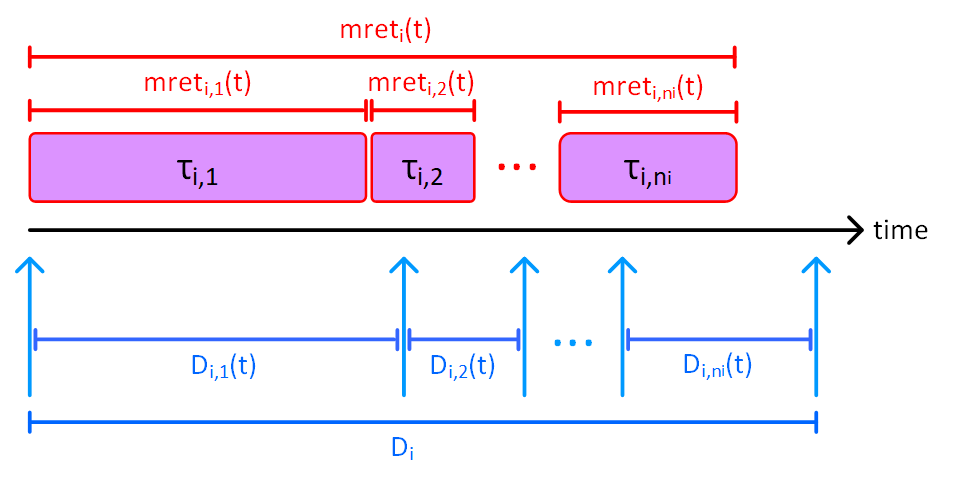}
        \caption{\label{fig:task}Task staging and virtual deadline assignment}
        \vspace{-15pt}
    \end{figure}

    Each task is represented as $\tau_i(T_i, D_i, mret_i(t), p_i, ctx_i(t))$, where $T_i$ is the period, $D_i$ the relative deadline, $mret_i(t)$ the Maximum Recent Execution Time (MRET) at time $t$, $p_i$ the priority level, and $1 \leq ctx_i(t) \leq N_c$ the current context. Deadlines are set equal to periods ($D_i=T_i$). MRET is used instead of WCET to avoid pessimistic estimates common with colocated DNNs (see \cref{sec:mret}). A stage is defined as $\tau_{i,j}(D_{i,j}(t), mret_{i,j}(t))$, where $D_{i,j}$ is the virtual deadline (determined using \cref{eq:deadline}). \cref{fig:task} shows task stages and their virtual relative deadlines based on MRET.

    \subsection{\label{sec:defs}Definitions}
    \subsubsection{\label{sec:staging}Staging}
    To be able to enforce priorities more efficiently on a smaller scale, we introduce a synchronization-based, coarse-grained preemption mechanism called \textit{staging}. Synchronization points can be placed after a few kernels \cite{han2022microsecond}, after each layer \cite{bai2020pipeswitch}, after several layers \cite{zhang2023shepherd}, or even dynamically \cite{xia2023towards}. Excessive synchronization, however, can reduce GPU utilization, as deep learning frameworks release kernels asynchronously to optimize throughput. We segment DNNs into sequential stages/sub-tasks, allowing preemption only at these boundaries. Stages are selected based on the DNN's logical structure; for instance, ResNet \cite{he2016deep} is divided into four stages.
    
    \subsubsection{\label{sec:mret}Maximum Recent Execution Time (MRET)}
    We propose a dynamic WCET estimation, \textit{Maximum Recent Execution Time} (MRET), capturing the maximum execution time within a recent window, adapting to workload changes. MRET is computed per stage rather than for the entire task (\cref{fig:task}).
    \begin{align}\label{eq:mret}
        mret_{i,j}(t) &= \max_{p \in \{t - ws, \ldots, t - 1\}} et_{i,j}(p)\\
        mret_i(t) &= \sum_{1 \leq j \leq n_i} mret_{i,j}(t),
    \end{align}
    Here, $ws$ is the window size; $mret_{i,j}$ and $et_{i,j}(t)$ are the MRET and actual execution time of $\tau_{i,j}$ at time $t$.
    
    \subsubsection{\label{sec:util}Utilization}
    In DARIS, we define the utilization of a task as:
    \begin{equation}\label{eq:utask}
        u_i(t) = \frac{mret_i(t)}{T_i}
    \end{equation}
    We use \Crefrange{eq:uhigh}{eq:utotal} to measure the utilization of a context:
    \begin{align}
        U^{h,t}_k(t) &= \sum_{\substack{1 \leq i \leq N_h \\ ctx_i=k}} u_i^h(t)\label{eq:uhigh}\\
        U^{l,t}_k(t) &= \sum_{\substack{1 \leq i \leq N_l \\ ctx_i=k}} u_i^l(t)\label{eq:ulow}\\
        U^t_k(t) &= U^{h,t}_k(t)+U^{l,t}_k(t)\label{eq:utotal}
    \end{align}
    where $U^t_k(t)$ is the total utilization of context $k$, and $U^{h,t}_k(t)$ and $U^{l,t}_k(t)$ are the total utilization of HP and LP tasks in context $k$, respectively. We will use \cref{eq:utotal} for load balancing between contexts in the offline phase. For the admission test during the online phase, we will rely on the active utilization:
    \begin{equation}\label{eq:uactive}
        U^a_k(t)=U^{h,t}_k(t)+U^{l,a}_k(t)
    \end{equation}
    The active utilization is defined as the sum of the total utilization of the HP tasks ($U^{h,t}_k(t)$) and the total utilization of \textit{currently active} LP tasks ($U^{l,a}_k(t)$), i.e., LP tasks that have an active job released but not yet finished.
    
    \begin{figure}
        \centering
        \includegraphics[width=0.7 \linewidth]{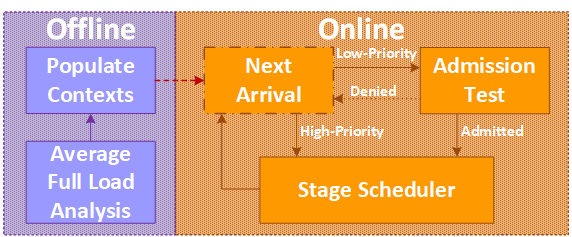}
        \caption{\label{fig:scheduler}Proposed Scheduler}
        \vspace{-15pt}
    \end{figure}

    \subsubsection{\label{sec:deadline}Virtual Deadline}
    The task deadlines are usually an application requirement set by the designer. We introduce the \textit{virtual deadline}, an intermediary parameter assigned to each stage. The virtual deadline for $\tau_{i,j}$ is calculated as:
    \begin{equation}\label{eq:deadline}
        D_{i,j}(t) = \frac{mret_{i,j}(t)}{mret_i(t)} D_i
    \end{equation}
    \cref{fig:task} illustrates the relationship between MRET and virtual deadlines, with longer stages receiving a larger share of the task's total relative deadline ($D_i$).

    \subsection{\label{sec:params}Hyperparameters}
    \subsubsection{\label{sec:count}Number of Parallel Tasks}
    Each stream can execute one task at a time, with the total number of streams determining the maximum number of jobs. We use $N_c$ contexts, each containing $N_s$ streams, allowing a maximum of $N_p = N_c \times N_s$ concurrent DNNs. The hyperparameters $N_c$ and $N_s$ define the desired parallel task count and need to be determined. Three policies, based on the context and stream numbers, are discussed in \cref{sec:impl}.
     
    \subsubsection{\label{sec:sm}Oversubscription Level and Number of SMs}
    All contexts are allocated an equal number of SMs as:
    \begin{equation}
        N_{SM} = \text{ceil}_{\text{even}}\left(\frac{OS \times N_{SM,max}}{N_c}\right)
    \end{equation}
    where $\text{ceil}_{\text{even}}$ rounds up to the nearest even number, $N_{SM,max}$ is the total GPU SMs, and $OS$ is the oversubscription value, constrained to $1 \leq OS \leq N_c$. $OS=1$ assigns each context its own SMs, with $OS=N_c$ context will share all SMs, anything in between has a variable level of SM sharing (i.e. lower values reduces interference while higher values enhance utilization through increased sharing).

    \section{\label{sec:scheduler}Real-Time Scheduler}
    We propose a multi-tenant inference scheduler for real-time DNNs, where each task corresponds to a distinct DNN and is classified as either HP or LP. Our scheduler (\cref{fig:scheduler}) has two phases: offline and online. The offline phase sets the initial state, while the online phase is the main body of the scheduler.

    \subsection{\label{sec:offline}Offline Phase}
    In the offline phase, we allocate $N_c$ contexts to tasks. HP tasks are given fixed contexts, while LP tasks can migrate between contexts as needed. This initial assignment prioritizes load balancing across contexts, establishing an efficient starting point for the online phase.

    \subsubsection{\label{sec:worst}Average Full Load Analysis}
    In the offline phase, with no measurement history, MRET cannot be used. Instead, we compute the \textit{Average Full-Load Execution Time} (AFET) by executing the target task in one stream while randomly executing others in the remaining streams, providing a pessimistic initialization metric. AFET is replaced in later iterations as it does not adapt to recent load trends (existing WCET estimates can be used if desired). Substituting $\mathit{afet_i}$ for $mret_i(t)$ in \cref{eq:utask} when $t=0$, task utilization is computed as:
    \begin{equation}
         u_i(t)=
        \begin{cases}
            \frac{afet_i}{T_i} & \text{if } t = 0 \\
            \frac{mret_i(t)}{T_i} & \text{if } t \geq 1
        \end{cases}
    \end{equation}

    \begin{table}
        \centering
        \caption{Batching performance of different DNNs}
        \label{tab:batch}
        \begin{tabular}{c|c|c|c}
             DNN & min (JPS) & max (JPS) & batching gain\\
             \hline
             ResNet18 & 627 & 1025 & 1.63x\\
             ResNet50 & 250 & 433 & 1.73x\\
             UNet & 241 & 260 & 1.08x\\
             InceptionV3 & 142 & 446 & 3.13x
        \end{tabular}
    \end{table}
    
    \begin{algorithm}
        \caption{\label{alg:init_context}Initial Context Assignment}
        \begin{algorithmic}[1]
            \Procedure{PopulateContexts}{ }
            \State // $pool: \text{context pool}$
            \ForAll{$task$ \textbf{in} $highTasks$} \label{line:3}
                \State $ctx \gets minUtil(pool)$
                \State $task.context \gets ctx$
                \State $ctx.totalUtil \gets ctx.totalUtil + task.util$
            \EndFor \label{line:7}

            \ForAll{$task$ \textbf{in} $lowTasks$} \label{line:8}
                \State $ctx \gets minUtil(pool)$
                \State $task.context \gets ctx$
                \State $ctx.totalUtil \gets ctx.totalUtil + task.util$
            \EndFor \label{line:12}
        \EndProcedure
        \end{algorithmic}
    \end{algorithm}
    
    \subsubsection{\label{sec:init_context}Populating Contexts}
    The objective is to assign tasks to contexts such that total context utilization ($U^t_k(0)$), HP ($U^{h,t}_k(0)$), and LP ($U^{l,t}_k(0)$) task utilization are balanced across contexts. \cref{alg:init_context} presents the pseudo-code for the initial context assignment. \Crefrange{line:3}{line:7} assigns HP tasks to contexts, while \Crefrange{line:8}{line:12} distributes LP tasks to balance utilization.

    \subsection{\label{sec:online}Online Phase}
    The goal of the online phase is to promptly schedule HP tasks, minimize deadline misses, and maximize throughput. This phase consists of two main components, discussed in the following sections.

    \subsubsection{\label{sec:test}Admission Test}
    A utilization-based admission mechanism for LP tasks is preferable when demand exceeds capacity. LP tasks undergo a utilization-based test within each context. HP tasks reserve a portion of context utilization, leaving the remaining capacity for LP tasks, which are subject to this utilization test. Given multiple streams ($N_s$) within each context, the utilization test for $ctx_k$ at time $t$ is defined as:
    \begin{align}
        &U^r_k(t) = N_s - U^{h,t}_k(t)\\
        &U^{l,a}_k(t) + u_j(t) <  U^r_k(t)\label{eq:test}
    \end{align}    
    Here, $U^r_k(t)$ is the remaining utilization of context $k$. If a job satisfies \cref{eq:test} for $k=ctx_i(t)$, it will be scheduled in context $k$. Otherwise, we will test other contexts ($k \in \{1, \dots, N_c\} \setminus \{ctx_i(t)\}$) as migrations candidates. If any of them satisfy \cref{eq:test}, $\tau_i$ will be migrated to the context with earliest predicted finish time. If no context is found that passes the admission test, the task is rejected.

    \subsubsection{\label{sec:opsched}Stage Scheduler}
    We extend task priorities from two to eight fixed levels for stages. Stages from HP tasks always take precedence over LP tasks, with the last stage of each task ($\tau_{i,n_i}$) prioritized higher. Any stage with an immediate preceding missed virtual deadline has the next priority level (e.g., if $\tau_{i,j}$ misses its deadline, $\tau_{i,j+1}$ receives higher priority). This hierarchy ensures HP tasks are serviced first, emphasizes the final stage of each task to prevent overall deadline misses, and prevents cascading misses by prioritizing stages with preceding deadline misses. Finally, EDF within fixed priority levels ensures tasks with approaching deadlines are prioritized.
    
    \begin{table}
        \centering
        \caption{Task sets}
        \label{tab:tasks}
        \begin{tabular}{c|c|c|c}
             Name & \#High & \#Low & Task JPS\\
             \hline
             ResNet18 & 17 & 34 & 30\\
             UNet & 5 & 10 & 24\\
             InceptionV3 & 9 & 18 & 24
        \end{tabular}
    \end{table}
    
    \begin{figure*}
        \centering
        \begin{subfigure}{0.49\linewidth}
            \includegraphics[width=1 \linewidth]{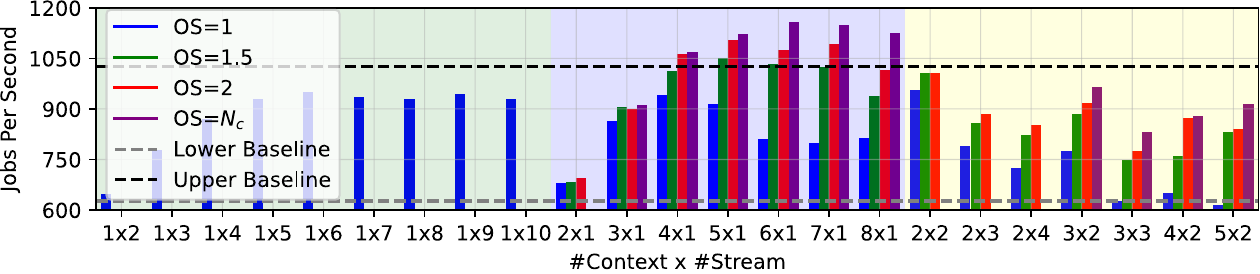}
            \vspace{-15pt}
           \caption{\label{fig:res_jps}Throughput}
        \end{subfigure}
        \centering
        \begin{subfigure}{0.49\linewidth}
            \includegraphics[width=1 \linewidth]{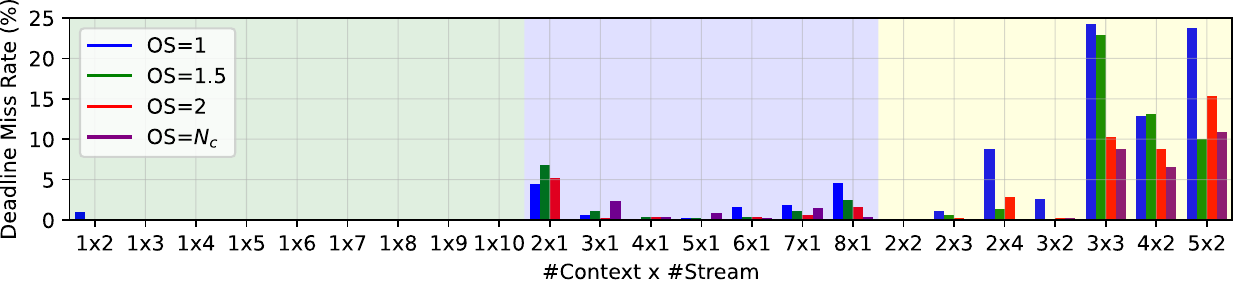}
            \vspace{-15pt}
            \caption{\label{fig:res_dmr}LP Deadline Misses}
        \end{subfigure}
        \vspace{-5pt}
        \caption{\label{fig:res}Scheduling results for ResNet18 task set (\colorbox{green!15}{STR}, \colorbox{blue!15}{MPS}, \colorbox{yellow!20}{MPS+STR})}
        \vspace{-10pt}
    \end{figure*}

    \begin{figure*}
        \centering
        \begin{subfigure}{0.49\linewidth}
            \includegraphics[width=1 \linewidth]{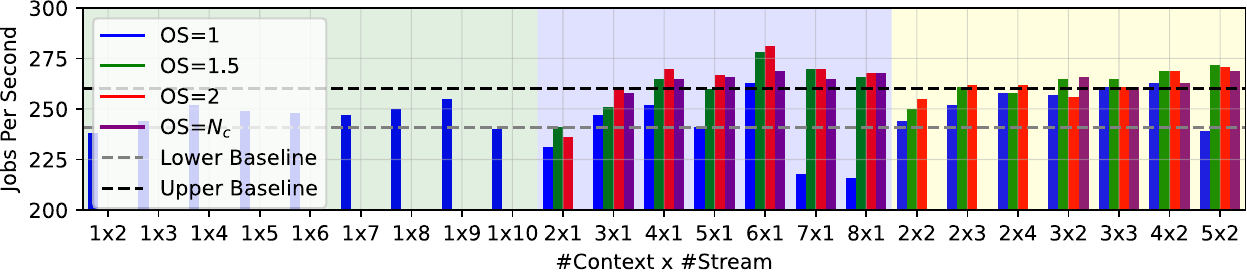}
            \vspace{-15pt}
            \caption{\label{fig:unt_jps}Throughput}
        \end{subfigure}
        \begin{subfigure}{0.49\linewidth}
            \includegraphics[width=1 \linewidth]{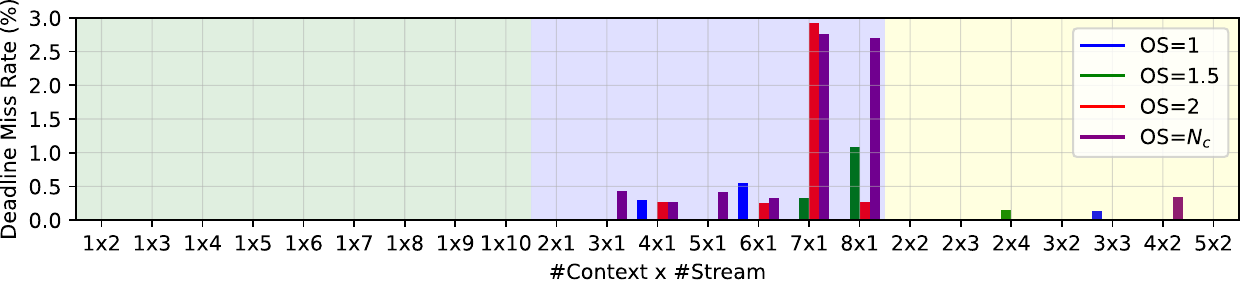}
            \vspace{-15pt}
            \caption{\label{fig:unt_dmr}LP Deadline Misses}
        \end{subfigure}
        \vspace{-5pt}
        \caption{\label{fig:unt}Scheduling results for UNet task set (\colorbox{green!15}{STR}, \colorbox{blue!15}{MPS}, \colorbox{yellow!20}{MPS+STR})}
        \vspace{-10pt}
    \end{figure*}

    \begin{figure*}
        \centering
        \begin{subfigure}{0.49\linewidth}
            \includegraphics[width=1 \linewidth]{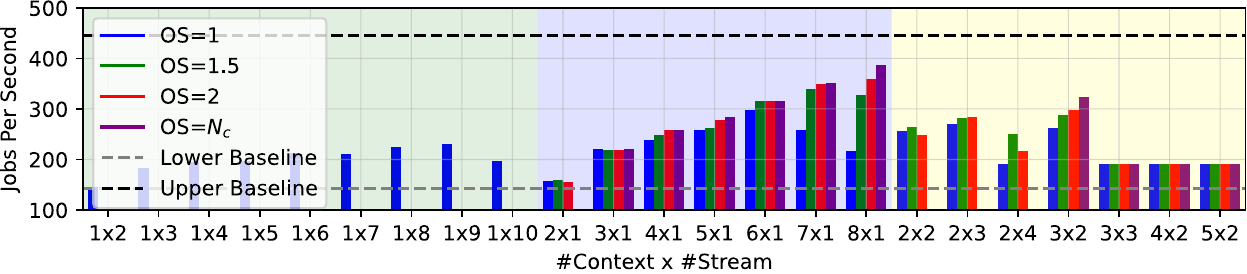}
            \vspace{-15pt}
            \caption{\label{fig:inc_jps}Throughput}
        \end{subfigure}
        \begin{subfigure}{0.49\linewidth}
            \includegraphics[width=1 \linewidth]{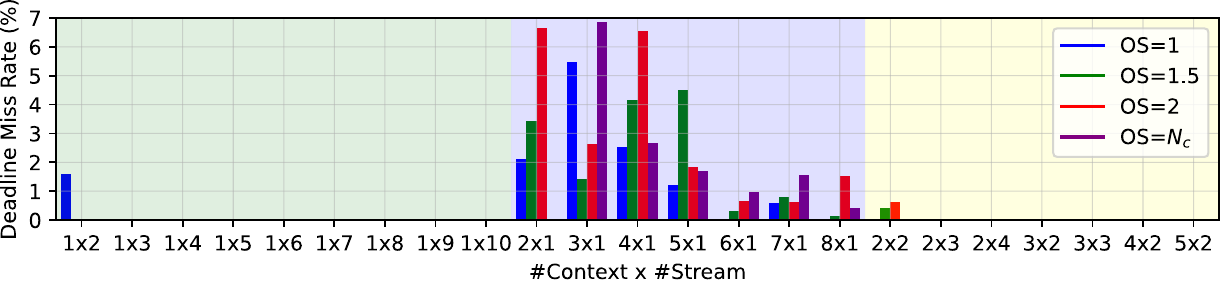}
            \vspace{-15pt}
            \caption{\label{fig:inc_dmr}LP Deadline Misses}
        \end{subfigure}
        \vspace{-5pt}
        \caption{\label{fig:inc}Scheduling results for InceptionV3 (\colorbox{green!15}{STR}, \colorbox{blue!15}{MPS}, \colorbox{yellow!20}{MPS+STR})}
        \vspace{-10pt}
    \end{figure*}

    \section{\label{sec:impl}Implementation and Setup}    
    We evaluated DARIS on an RTX 2080 Ti GPU implemented using LibTorch (PyTorch’s C++ API). Backend modifications enabled multi-context support via thread-local variables. PyTorch’s multi-device API was customized to treat each device as a context while replacing device-level with context-level synchronizations to avoid inter-context locks. Key components for streams, memory, cache, and cuBLAS were updated to support our algorithm.
    
    Our benchmarks include three DNNs: \textit{ResNet} \cite{he2016deep}, \textit{UNet} \cite{ronneberger2015u}, and \textit{InceptionV3} \cite{szegedy2016rethinking}, each with 224x224x3 input. Throughput gains from batching, evaluated as an upper limit for throughput without colocation, are reported in \cref{fig:batch} and \cref{tab:batch} (throughput in \textit{Jobs Per Second} (JPS)). ResNet, with a linear design, is widely used, while UNet’s wide architecture and skip connections make it memory-heavy, achieving only a 1.08x from batching. InceptionV3, with multiple parallel paths, exhibits the highest batching gain of 3.13x, highlighting its reliance on batching.

    We use MPS and CUDA Streams with oversubscription to colocate DNNs, aiming to minimize deadline misses and maximize throughput. DARIS variations are assessed using three partitioning policies:
    \begin{enumerate}
        \item \textbf{\textit{STR}}: Exclusively uses streams for scheduling DNNs.
        \item \textbf{\textit{MPS}}: Relies solely on MPS for scheduling DNNs.
        \item \textbf{\textit{MPS+STR}}: Combines MPS and streams to assess their joint effect on timeliness and throughput.
    \end{enumerate}
    We schedule 2–10 parallel DNNs ($2 \leq N_p \leq 10$). Four oversubscription options are explored: $OS \in \{1, 1.5, 2, N_c\}$, where $OS=1$ indicates no SM sharing and $OS=N_c$ represents full SM sharing.

    Performance is assessed using total JPS and Deadline Miss Rate (DMR) for throughput and timeliness, with configurations denoted as $N_c \times N_s$ or $N_c \times N_s\_OS$. Three main task sets (\cref{tab:tasks}), each tied to a DNN type, are tested with 30 jobs/sec for ResNet18 and 24 jobs/sec for others. Experiments are run under 150\% overload, using the upper baseline as full load due to varying maximum loads across different configurations. A 2:1 LP-to-HP task ratio is maintained, with alternative ratios explored in \cref{sec:ratio}.

    \section{\label{sec:result}Experimental Results}
    Main scenario results are presented in \Crefrange{fig:res}{fig:inc}. Throughput figures (\Crefrange{fig:res_jps}{fig:inc_jps}) compare with lower (single DNN) and upper (pure batching) baseline throughput from \cref{tab:batch}. Although meeting every deadline is not guaranteed, we did not observe any HP deadline misses, and only LP DMRs are shown in \Crefrange{fig:res_dmr}{fig:inc_dmr}. DMR is the ratio of missed deadlines to accepted jobs.
    
    \subsubsection{\label{sec:throughput}Throughput}
    As shown in \Crefrange{fig:res_jps}{fig:inc_jps}, \textit{MPS} delivers the highest throughput across all DNNs. ResNet18 and UNet achieve peak throughput at $N_c=6$, while InceptionV3 benefits from increased concurrency up to $N_c=8$. The \textit{MPS+STR} policy outperforms \textit{STR}, showing that multiple contexts enhance throughput on MPS-enabled GPUs. For ResNet18 and UNet, we exceeded the pure batching baseline (\cref{tab:batch}) by 13\% (1158 JPS vs. 1025 JPS) and 8\% (281 JPS vs. 260 JPS) respectively, without batching. However, for InceptionV3, we only achieved 87\% of its upper baseline. Its complex, narrow architecture limits throughput. We conducted an experience to release parallel paths of InceptionV3 using distinct streams to improve throughput but gained only 9\% in throughput still below the upper baseline. By using batching instead, we were able to surpass upper baseline for InceptionV3 which will be discussed in \cref{sec:batching}.

    \subsection{\label{sec:timeliness}Deadline Misses}
    DARIS prioritizes minimizing HP task response time to reduce the chance of deadline misses, though it does not guarantee meeting every deadline. In our main experiments, no HP deadline misses were observed. From here on, DMR refers to LP DMR unless stated otherwise. Except for less than 2\% DMR in the $1 \times 2$ configuration for InceptionV3, no deadline misses occurred with the \textit{STR} policy. Each context uses a dedicated job queue, leaving \textit{STR} with a single global queue. While this may reduce throughput, it ensures optimal timeliness. UNet showed the lowest DMR, peaking at less than 3\%, with only 0.25\% at its best throughput configuration ($6 \times 1\_2$). The \textit{MPS+STR} policy produced the worst DMR, reaching 25\% for ResNet18. Meanwhile, with the \textit{MPS} policy, both ResNet18 and InceptionV3 maintained DMRs below 7\%, with around 2\% at their peak throughput configurations ($6 \times 1\_6$ and $8 \times 1\_8$, respectively).

    \subsection{Comparison with State-of-the-Art}
    GSlice \cite{dhakal2020gslice} is a state-of-the-art inference server, achieving 1152 JPS for ResNet50 with batching and 1193 JPS with GSlice—a 3.5\% gain. In comparison, we achieved 433 JPS with batching and 498 JPS using DARIS on our hardware, improving throughput by 15\% over batching and 11.5\% over GSlice. Without oversubscription, DARIS throughput drops to 374 JPS, 8\% below batching. Inference servers like \cite{dhakal2020gslice, gujarati2020serving} generally allow some deadline misses without detailed rates. The closest comparison, \cite{wang2021balancing}, reports no HP deadline misses and up to 12\% for LP tasks. In our case, LP deadline misses are below 2\% in the best and under 7\% in the worst scenario with the \textit{MPS} policy, and zero with the \textit{STR} policy. RTGPU \cite{zou2023rtgpu} lacks task prioritization, reporting up to 11\% overall deadline misses.

    \subsection{DARIS Policy Comparison}
    Overall observations indicate that the \textit{MPS} policy offers the best throughput, while \textit{STR} results in ideal DMR. Although \textit{MPS} may not achieve the lowest DMR, it remains below 7\% in all configurations. For many applications, this DMR is acceptable, but for those with stringent constraints, \textit{STR} is the safest policy. In scenarios with embedded GPUs lacking MPS support, \textit{STR} is the sole feasible option. Conversely, the \textit{MPS+STR} policy yields the least favorable outcomes, with subpar throughput and DMR across all configurations. Also, UNet has the most consistent performance in all scenarios, making it the least sensitive DNN to concurrency mechanisms and configurations.

    \subsection{Mixed Task Set}
\begin{figure}
        \centering
        \begin{subfigure}{0.49\linewidth}
           \includegraphics[width=1 \linewidth]{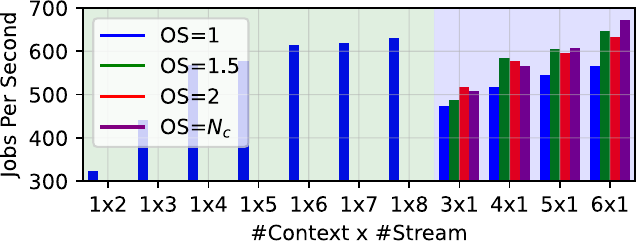}
           \caption{\label{fig:mix_jps}Throughput}
        \end{subfigure}
        \begin{subfigure}{0.49\linewidth}
           \includegraphics[width=1 \linewidth]{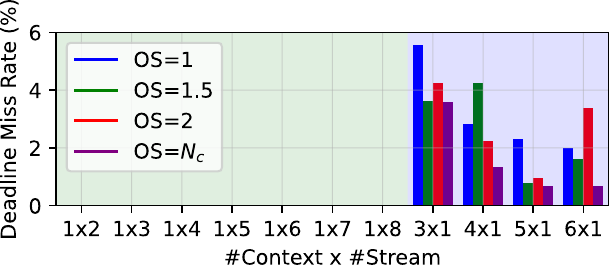}
           \caption{\label{fig:mix_dmr}LP Deadline Misses}
        \end{subfigure}
        \caption{\label{fig:mix}Scheduling results for mixed task set (\colorbox{green!15}{STR}, \colorbox{blue!15}{MPS})}
        \vspace{-10pt}
    \end{figure}

    It is essential to analyze a mixed task set with all DNN types, reflecting real-world scenarios. The results for the mixed task set are shown in \cref{fig:mix}. As with individual task sets, the \textit{MPS} policy achieves the highest throughput, while the \textit{STR} policy offers the most reliable deadline performance.

    \subsection{\label{sec:oversubscribe}Oversubscribing SMs}
    Contrary to some claims in the literature \cite{dhakal2022slice, dhakal2023d, rattihalli2023fine}, our study shows that, in real-time scenarios without batching, oversubscription significantly boosts throughput. While its impact on timeliness is not strictly monotonic, higher $OS$ values generally result in fewer deadline misses. Isolating SMs ($OS=1$) leads to a sharp drop in throughput. Our findings confirm that oversubscription consistently benefits both throughput and timeliness. For wide DNNs like UNet, which gain little from batching, 200\% SM oversubscription is sufficient, while narrower DNNs require even higher oversubscription. Although more concurrent DNNs combined with oversubscription can increase resource contention, it also enhances throughput. The goal is to identify a trade-off point to maximize benefit.

    \begin{figure}
        \centering
        \begin{subfigure}{0.45\linewidth}
           \includegraphics[width=1 \linewidth]{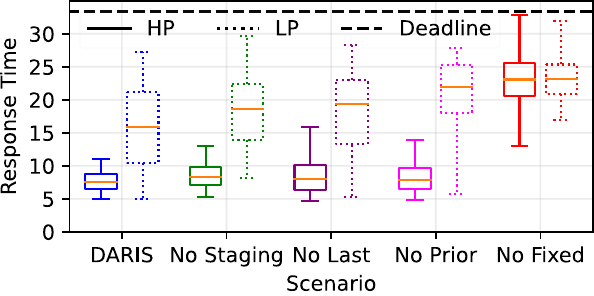}
           \caption{\label{fig:scn_rsp}Response time}
        \end{subfigure}
        \begin{subfigure}{0.45\linewidth}
           \includegraphics[width=1 \linewidth]{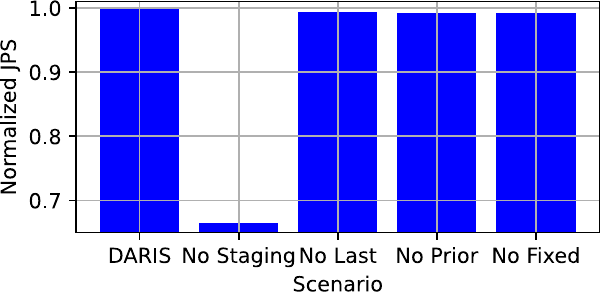}
           \caption{\label{fig:scn_jps}Normalized Throughput}
        \end{subfigure}        
        \caption{\label{fig:scn}Response time and throughput for different scenarios}
        \vspace{-15pt}
    \end{figure}

    \subsection{DARIS Modules Contribution}
    Works addressing real-time constraints often treat meeting deadlines or minimizing latency as binary outcomes \cite{dhakal2020gslice, wang2021balancing, zou2023rtgpu, gujarati2020serving}. However, enhancing the quality of service by minimizing response time is crucial for better responsiveness. We conducted experiments to evaluate how different DARIS modules impact overall performance. \cref{fig:scn_rsp} shows the response times, and \cref{fig:scn_jps} presents normalized throughput for DARIS and four alternative scenarios using ResNet18:
    \begin{itemize}
        \item \textit{No Staging}: tasks treated as whole units without staging
        \item \textit{No Last}: last stages of tasks are not prioritized
        \item \textit{No Prior}: no high priority when preceding stage misses deadline
        \item \textit{No Fixed}: no differentiation in task priority among stages
    \end{itemize}
    The original DARIS achieves response times of 5–12 ms for HP tasks and 5–27.5 ms for LP tasks, making HP tasks finish roughly 2.5 times faster. In \textit{No Staging} scenario, throughput drops by 33\%, and response times increase due to the lack of preemption, with 5.5\% and 22.5\% deadline misses for HP and LP tasks, respectively. \textit{No Last} scenario increases worst-case response times for HP tasks by 38\% without significantly affecting throughput. In \textit{No Prior} scenario, average response times rise for all tasks. \textit{No Fixed} scenario, which removes inter-task priority differentiation, results in a 2.5\% deadline miss rate for both task priorities. These findings highlight the effectiveness of DARIS modules, emphasizing the importance of \textit{staging} and \textit{task priority} in improving throughput and meeting deadlines.
    
    \begin{figure}
        \begin{center}
           \includegraphics[width=0.85 \linewidth]{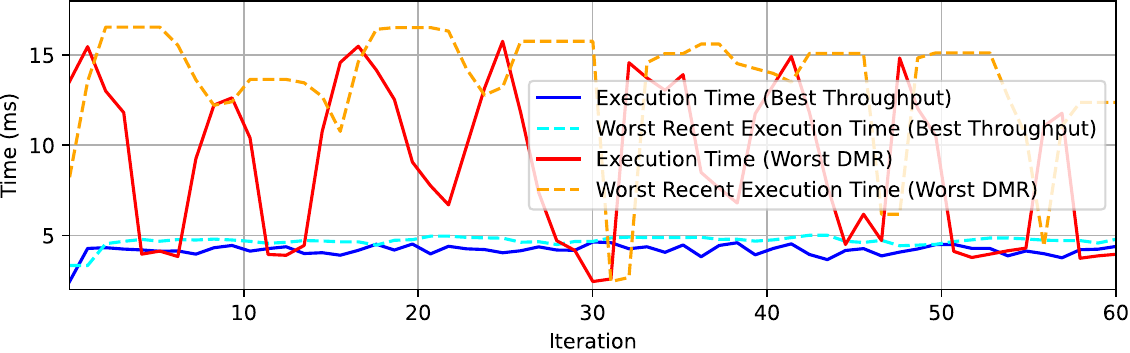}
        \end{center}
        \vspace{-10pt}
        \caption{\label{fig:res_wet}Execution time and MRET of ResNet18}
        \vspace{-15pt}
    \end{figure}

    \subsection{Maximum Recent Execution Time}
    We introduced MRET to address the unpredictability of GPU execution time and avoid the pessimism of history-based WCET approaches. \cref{fig:res_wet} shows the actual execution time and MRET for ResNet18 under the best throughput configuration ($6 \times 1\_6$) and the worst deadline miss rate configuration ($3 \times 3\_1$). We selected a window size ($ws$) of 5, as smaller values increase DMR, while larger values reduce throughput. With the $6 \times 1\_6$ configuration, MRET accurately predicts execution time in most cases, whereas in the $3 \times 3\_1$ configuration, execution time often exceeds MRET predictions.

    \begin{figure*}
        \centering
        \begin{subfigure}{0.3\linewidth}
           \includegraphics[width=\linewidth]{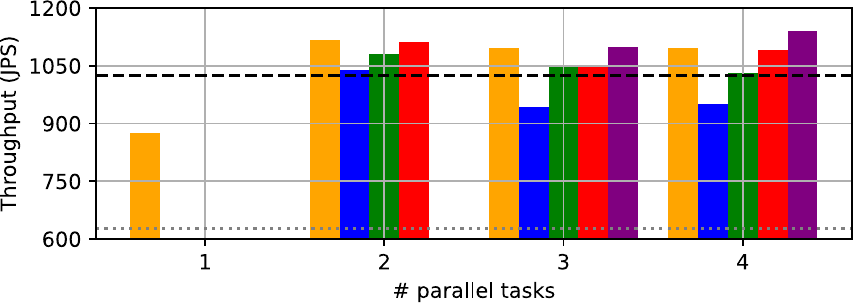}
           \caption{\label{fig:res_btc_jps}ResNet18 Batched Throughput}
        \end{subfigure}
        \begin{subfigure}{0.3\linewidth}
           \includegraphics[width=\linewidth]{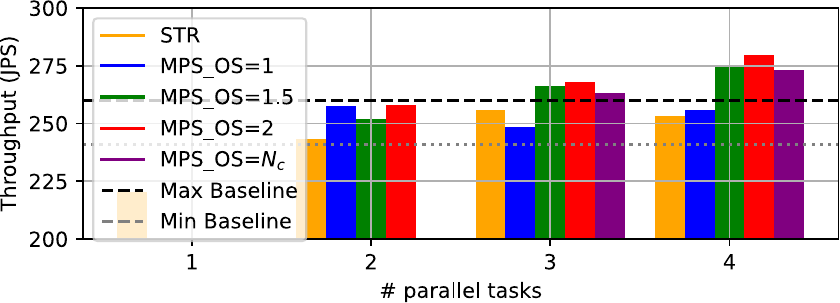}
           \caption{\label{fig:unt_btc_jps}UNet Batched Throughput}
        \end{subfigure}
        \begin{subfigure}{0.3\linewidth}
           \includegraphics[width=\linewidth]{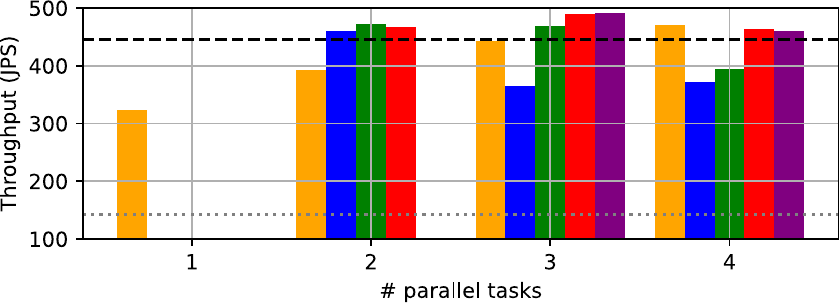}
           \caption{\label{fig:inc_btc_jps}InceptionV3 Batched Throughput}
        \end{subfigure}
        \begin{subfigure}{0.3\linewidth}
           \includegraphics[width=\linewidth]{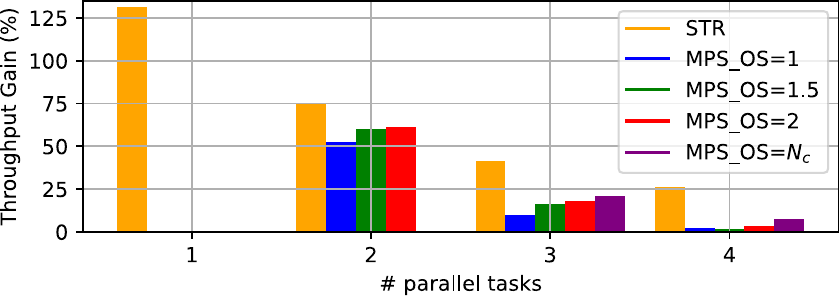}
           \caption{\label{fig:res_btc_jps_p}ResNet18 Batched Throughput Gain}
        \end{subfigure}
        \begin{subfigure}{0.3\linewidth}
           \includegraphics[width=\linewidth]{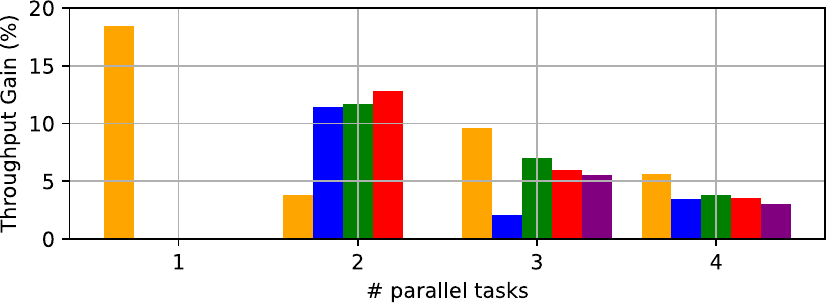}
           \caption{\label{fig:unt_btc_jps_p}UNet Batched Throughput Gain}
        \end{subfigure}
        \begin{subfigure}{0.3\linewidth}
           \includegraphics[width=\linewidth]{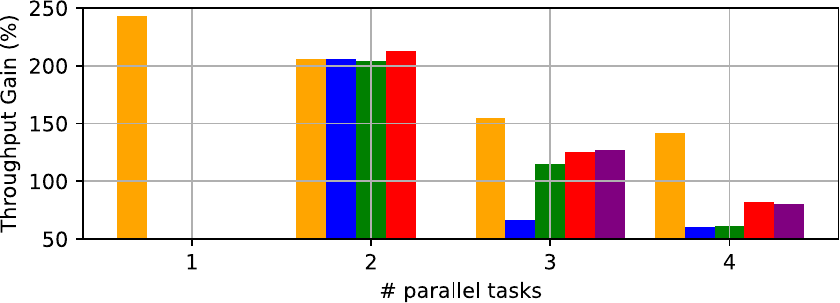}
           \caption{\label{fig:inc_btc_jps_p}InceptionV3 Batched Throughput Gain}
        \end{subfigure}
        \begin{subfigure}{0.3\linewidth}
           \includegraphics[width=\linewidth]{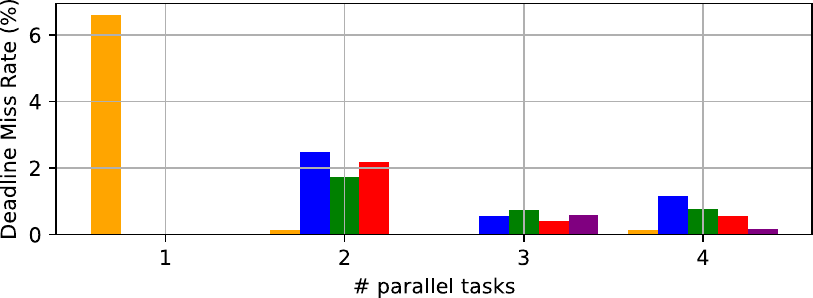}
           \caption{\label{fig:res_btc_dmr}ResNet18 LP DMR}
        \end{subfigure}
        \begin{subfigure}{0.3\linewidth}
           \includegraphics[width=\linewidth]{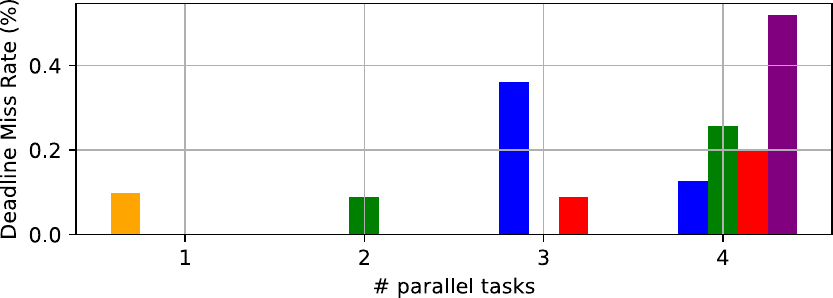}
           \caption{\label{fig:unt_btc_dmr}UNet LP DMR}
        \end{subfigure}
        \begin{subfigure}{0.3\linewidth}
           \includegraphics[width=\linewidth]{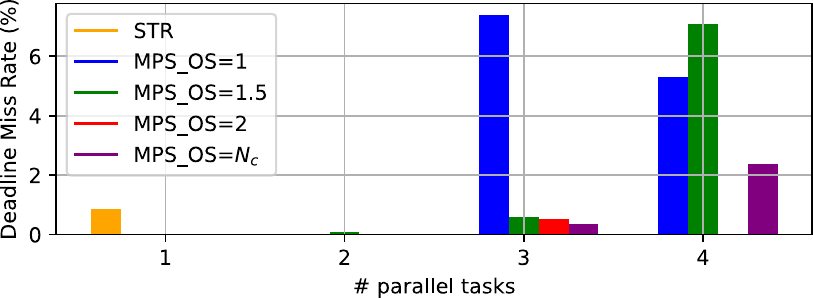}
           \caption{\label{fig:inc_btc_dmr}InceptionV3 LP DMR}
        \end{subfigure}        
        \caption{\label{fig:btc_jps}Absolute Throughput When Batching}
    \end{figure*}

    \begin{figure*}
        \centering
        \begin{subfigure}{0.33\linewidth}
           \includegraphics[width=\linewidth]{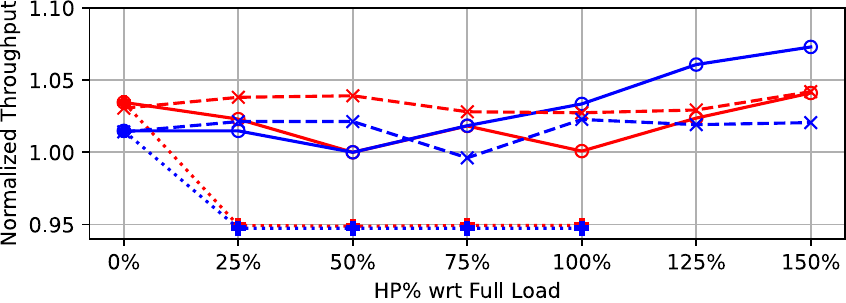}
           \caption{\label{fig:old_jps}Normalized Throughput}
        \end{subfigure}
        \centering
        \begin{subfigure}{0.285\linewidth}
           \includegraphics[width=\linewidth]{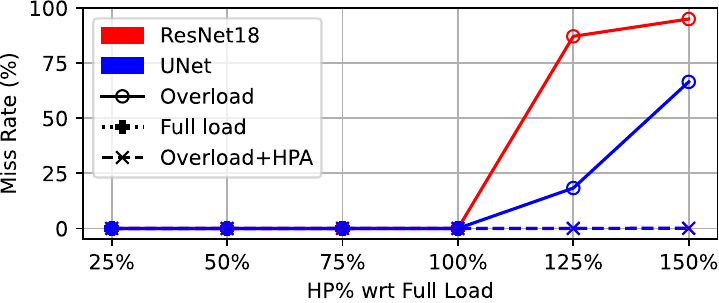}
           \caption{\label{fig:old_hmr}High-Priority DMR}
        \end{subfigure}
        \centering
        \begin{subfigure}{0.285\linewidth}
           \includegraphics[width=\linewidth]{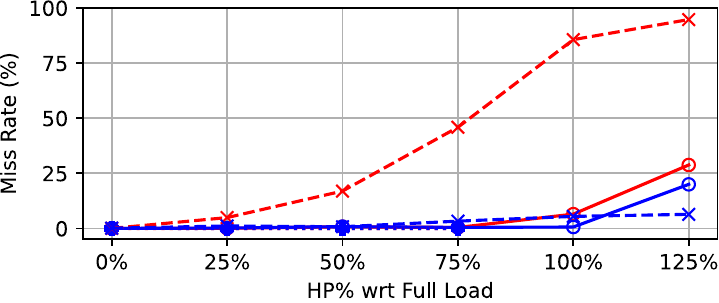}
           \caption{\label{fig:old_lmr}Low-Priority DMR}
        \end{subfigure}
        
        \caption{\label{fig:old}Overloading with different HP to LP ratios}
    \end{figure*}

    \subsection{\label{sec:batching}Batching}
    We conducted an additional experiment to demonstrate the impact of batching combined with DARIS, using batch sizes of 4, 2, and 8 for ResNet18, UNet, and InceptionV3, respectively. \Crefrange{fig:res_btc_jps}{fig:inc_btc_jps} present the throughput results. Our key observation is that fewer parallel tasks are needed to exceed the upper baseline, with decent throughput achieved even with 1 or 2 parallel tasks. We also observed similar benefits from SM oversubscription compared to the main experiment (\Crefrange{fig:res}{fig:inc}) without batching, though the difference is less pronounced.

    \Crefrange{fig:res_btc_jps_p}{fig:inc_btc_jps_p} show the throughput improvements compared to similar configurations from the main experiment (\Crefrange{fig:res_jps}{fig:inc_jps}). UNet exhibits the smallest gain, with improvements up to 18\%, reaffirming its efficient architecture in utilizing GPU resources even without batching. In contrast, InceptionV3 achieves the highest improvement, with at least a 55\% increase over the main experiment without batching. For DMR (\Crefrange{fig:res_btc_dmr}{fig:inc_btc_dmr}), ResNet18 and InceptionV3 show slightly better DMRs compared to the main experiment (\cref{fig:res_dmr}and \cref{fig:inc_dmr}). UNet demonstrates the most significant improvement, with its DMR reduced to under 0.5\%.

    \subsection{\label{sec:ratio}Overloading And Task Ratio}
    We conducted a final experiment to examine the scheduler's behavior under different HP-to-LP task ratios. \cref{fig:old} presents the results for throughput and deadline miss rates for both priorities, using ResNet18 and UNet in full load and overloaded scenarios. We also evaluate an overloaded scenario where HP tasks must pass the admission test (\textit{Overload+HPA}).

    As shown in \cref{fig:old_jps}, throughput remains stable across different task ratios and overload scenarios. In the full load scenario, throughput drops consistently by 5\% with the presence of LP tasks, but no HP or LP deadline misses occur. However, in the overload scenario, when HP load exceeds 100\% of full capacity, DMR for HP tasks rises sharply since HP tasks are scheduled without an admission test. Although this ensures all HP tasks are scheduled, it results in exponentially increasing deadline misses when the system is overwhelmed by HP tasks.
    
    To address this, we introduce \textit{Overload+HPA}, ensuring zero deadline misses for HP tasks by applying the admission test, even when the system is overloaded with HP tasks. The trade-off is that some HP tasks may be dropped, and DMR for LP tasks increases. However, UNet avoids this downside due to its wide efficient architecture. We recommend limiting HP tasks to 50\% of the full load to minimize deadline misses, as shown in the main experiments. In cases of overloaded HP tasks, using the \textit{Overload+HPA} approach ensures safer scheduling with minimal HP deadline misses.

    \section{\label{sec:conclusion}Conclusions}
    In this work, we presented DARIS, a novel real-time GPU scheduler designed for periodic tasks in a multi-tenant DNN setup with task priorities. DARIS applies oversubscribed spatio-temporal scheduling to minimize deadline misses for high-priority tasks while maximizing throughput by scheduling as many low-priority tasks as possible. For spatial partitioning, we used both MPS and CUDA streams to colocate multiple DNNs. Through extensive experiments, we compared the throughput and timeliness achieved by using MPS and streaming separately and in combination. We found that MPS provides the highest throughput, while streaming yields the lowest deadline miss rate. Additionally, we evaluated DARIS under batched inputs, overloaded scenarios, and various task ratios. Our experiments showed that oversubscribing SMs can boost throughput significantly, even in the presence of batched inputs and DNNs with wide architecture.

    For temporal scheduling, we used staging as synchronization points to enable coarse-grained preemption, improving throughput and reducing deadline misses. By prioritizing stages based on task priority, stage order, and past missed deadlines, we further reduced deadline misses and enhanced response times to improve the quality of service. While we explored leveraging the parallel paths of non-linear DNNs to boost throughput, we found that batched inputs provided superior results in managing parallel paths.

    \section*{Acknowledgment}
    This work was supported in part by NSF under grant number CPS-2038726 and by the Commonwealth Cyber Initiative, an investment in the advancement of cyber R\&D, innovation, and workforce development. For more information about CCI, visit \hyperlink{www.cyberinitiative.org}{www.cyberinitiative.org}.
    
    \printbibliography
    % \bibliography{references}
\end{document}